\documentclass{article}
\usepackage[T1]{fontenc}
\usepackage[utf8]{inputenc}
\usepackage{ismir,amsmath,cite,url}
\usepackage{graphicx}
\usepackage{color}
\usepackage{nicefrac}
\usepackage{amssymb}
\usepackage{amsmath}
\usepackage{booktabs}
\usepackage{multirow}
\usepackage[linesnumbered,ruled]{algorithm2e}
\usepackage{caption}
\usepackage{subcaption}

\newcommand{\mb}{\mathbf}
\newcommand{\dataset}{OLGA}

\usepackage[colorinlistoftodos,prependcaption,textsize=tiny]{todonotes}

\title{Artist Similarity with Graph Neural Networks}

\oneauthor
{Filip Korzeniowski \hspace{1cm} Sergio Oramas \hspace{1cm} Fabien Gouyon}
{Pandora Media LLC., Oakland, California, USA}

\def\authorname{Filip Korzeniowski, Sergio Oramas, and Fabien Gouyon}

\begin{document}

\maketitle

\begin{abstract}
Artist similarity plays an important role in organizing, understanding, and subsequently, facilitating discovery in large collections of music. In this paper, we present a hybrid approach to computing similarity between artists using graph neural networks trained with triplet loss. The novelty of using a graph neural network architecture is to combine the topology of a graph of artist connections with content features to embed artists into a vector space that encodes similarity.
To evaluate the proposed method, we compile the new \dataset{}~dataset, which contains artist similarities from AllMusic, together with content features from AcousticBrainz. With 17,673~artists, this is the largest academic artist similarity dataset that includes content-based features to date.
Moreover, we also showcase the scalability of our approach by experimenting with a much larger proprietary dataset.
Results show the superiority of the proposed approach over current state-of-the-art methods for music similarity.
Finally, we hope that the \dataset{}~dataset will facilitate research on data-driven models for artist similarity. 
\end{abstract}

\section{Introduction}

Music similarity has sparked interest early in the Music Information Retrieval community~\cite{aucouturier_musicsimilaritymeasures_2002,ellis_questgroundtruth_2002}, and has since then become a central concept for music discovery and recommendation in commercial music streaming services. 

There is however no consensual notion of \emph{ground-truth} for music similarity, as several viewpoints are relevant~\cite{ellis_questgroundtruth_2002}. 
For instance, music similarity can be considered at several levels of granularity; musical items of interest can be musical phrases, tracks, artists, genres, to name a few. Furthermore, the perception of similarity between two musical items can focus either on (1)~comparing \emph{descriptive} (or \emph{content-based}) aspects, such as the melody, harmony, timbre (in acoustic or symbolic form),  or (2)~\emph{relational} (sometimes called \emph{cultural}) aspects, such as listening patterns in user-item data, frequent co-occurrences of items in playlists, web pages, et cetera. 

In this paper, we focus on \emph{artist}-level similarity, and formulate the problem as a \emph{retrieval} task: given an artist, we want to retrieve the most similar artists, where the ground-truth for similarity is \emph{cultural}. More specifically, artist similarity is defined by music experts in some experiments, and by the ``wisdom of the crowd'' in other experiments. 

A variety of methods have been devised for computing artist similarity, from the use of audio descriptors to measure similarity~\cite{pohle_rhythmgeneralmusic_2009}, to leveraging text sources by measuring artist similarity as a document similarity task~\cite{schedl_harvestingmicroblogscontextual_2014}. A significant effort has been dedicated to the study of graphs that interconnect musical entities with semantic relations as a proxy to compute artist similarity. For instance, in~\cite{celma_foafingmusicbridging_2008}, user profiles, music descriptions and audio features are combined in a domain specific ontology to compute artist similarity, whereas in~\cite{oramas_semanticbasedapproachartist_2015}, semantic graphs of artists are extracted from artist biographies. 

Other approaches use deep neural networks to learn artist embeddings from heterogeneous data sources and then compute similarity in the resulting embedding space~\cite{mcfee_heterogeneousembeddingsubjective_2009}. 
More recently, metric learning approaches trained with triplet loss have been applied to learn the embedding space where similarity is computed~\cite{lee_disentangledmultidimensionalmetric_2020,doras_combiningmusicalfeatures_2020,lee_metriclearningvs_2020,park_representationlearningmusic_2018,yesiler_accuratescalableversion_2020,dorfer_learningaudiosheetmusic_2017}. 

In this work, we propose a novel artist similarity model that combines graph approaches and embedding approaches using graph neural networks. Our proposed model, described in details in Sec.~\ref{sec:modelling}, uses content-based features (audio descriptors, or musicological attributes) together with explicit similarity relations between artists made by human experts (or extracted from listener feedback). These relations are represented in a graph of artists; the topology of this graph thus
reflects the contextual aspects of artist similarity.

Our graph neural network is trained using triplet loss to learn a function that embeds artists using content features and graph connections. In this embedding space, similar artists are close to each other, while dissimilar ones are further apart.

To evaluate our approach (see Sec.~\ref{sec:experiments}), we compile a new dataset from publicly available sources, with similarity information and audio-based features for 17,673 artists, which we describe in Sec.~\ref{sec:datasets}.
In addition, we evaluate the scalability of our method using a larger, proprietary dataset with more than 136,731 artists.

\section{Modelling}\label{sec:modelling}

The goal of an artist similarity model is to define a
function \( s(a, b) \) that estimates the similarity of two artists---i.e.,
yields a large number if artist \( a \) is considered similar to artist \(b\),
and small number if not.

Many content-based methods for similarity estimation 
have
been developed
in the last decades of MIR research. The field has closely followed the
state-of-the-art in machine learning research, with general improvements coming
from the latter translating well into improvements in the former. Acknowledging
this fact, we select our baselines based on the most recent developments:
Siamese neural networks trained with variants of the triplet loss ~\cite{doras_combiningmusicalfeatures_2020,lee_metriclearningvs_2020,park_representationlearningmusic_2018,yesiler_accuratescalableversion_2020,dorfer_learningaudiosheetmusic_2017}. Building and training this type of models falls under the umbrella of \emph{metric learning.}

\subsection{Metric Learning}

The fundamental idea of metric learning is to learn a projection \(\mb{y}_v = f\left(\mb{x}_v\right)\)
of the input features \(\mb{x}_v\) of an item \(v\) into a new vector space; this vector
space should be structured in a way such that the distances between points reflect the 
task at hand. In our case, we want similar artists to be close together in this space, 
and dissimilar artists far away.

There is an abundance of methods that embed items into a vector space, many rooted
in statistics, that have been applied to music similarity~\cite{slaney_learningmetricmusic_2008}.
In this paper, we use a neural network for this purpose. The idea of using neural
networks to embed similar items close to each other in an embedding space was pioneered
by~\cite{bromley_signatureverificationusing_1993}, with several improvements developed
in the following decades. Most notably, the contrastive learning objective---where two
items are compared to each other as a training signal---was replaced by the \emph{triplet
loss}\cite{hoffer_deepmetriclearning_2015,wang_learningfinegrainedimage_2014a}. 
Here, we observe three items simultaneously: the \emph{anchor} item \(\mb{x}_a\) is
compared to a \emph{positive} sample \(\mb{x}_p\) and a \emph{negative} sample \(\mb{x}_n\).
With the following loss formulation, the network is trained to pull the positive close to the anchor, while pushing
the negative further away from it:
\[
    \mathcal{L}\left(t\right) = 
        \Big[ 
            d\left( \mb{y}_a, \mb{y}_n \right)
            - d\left( \mb{y}_a, \mb{y}_p \right)
            + \Delta
        \Big]^{+},
\]
where \(t\) denotes the triplet \( (\mb{y}_a, \mb{y}_p, \mb{y}_n) \), \(d\left(\cdot\right)\) is a distance function (usually Euclidean or cosine), \(\Delta\) is the maximum margin enforced by the loss, and \([\cdot]^+\) is the ramp function.

As mentioned before, state-of-the-art music similarity models are almost exclusively based 
on learning deep neural networks using the triplet loss. We thus adopt this method as our 
baseline model, which will serve as a comparison point to the graph neural network we
propose in the following sections.

\subsection{Graph Neural Networks}

A set of artists and their known similarity relations can be seen as a graph, where
the artists represent the nodes, and the similarity relations their (undirected)
connections. Graph methods thus naturally lend themselves to model the artist
similarity problem~\cite{oramas_semanticbasedapproachartist_2015}. A particular set of graph-based models that has been gaining traction
recently are \emph{graph neural networks} (GNNs), specifically \emph{convolutional}
GNNs. Pioneered by~\cite{bruna_spectralnetworkslocally_2014}, convolutional GNNs
have become increasingly popular for modelling different tasks that can be
interpreted as graphs. We refer the interested reader
to~\cite{wu_comprehensivesurveygraph_2021} for a comprehensive and historical
overview of GNNs. For brevity, we will focus on the one specific model
our work is based on---the GraphSAGE model introduced
by~\cite{hamilton_inductiverepresentationlearning_2017} and refined
by~\cite{ying_graphconvolutionalneural_2018}---and use the term GNNs for
convolutional GNNs.

\subsubsection{Model Overview}\label{sec:model_overview}

\begin{figure}
 \centerline{
    \includegraphics[width=\columnwidth]{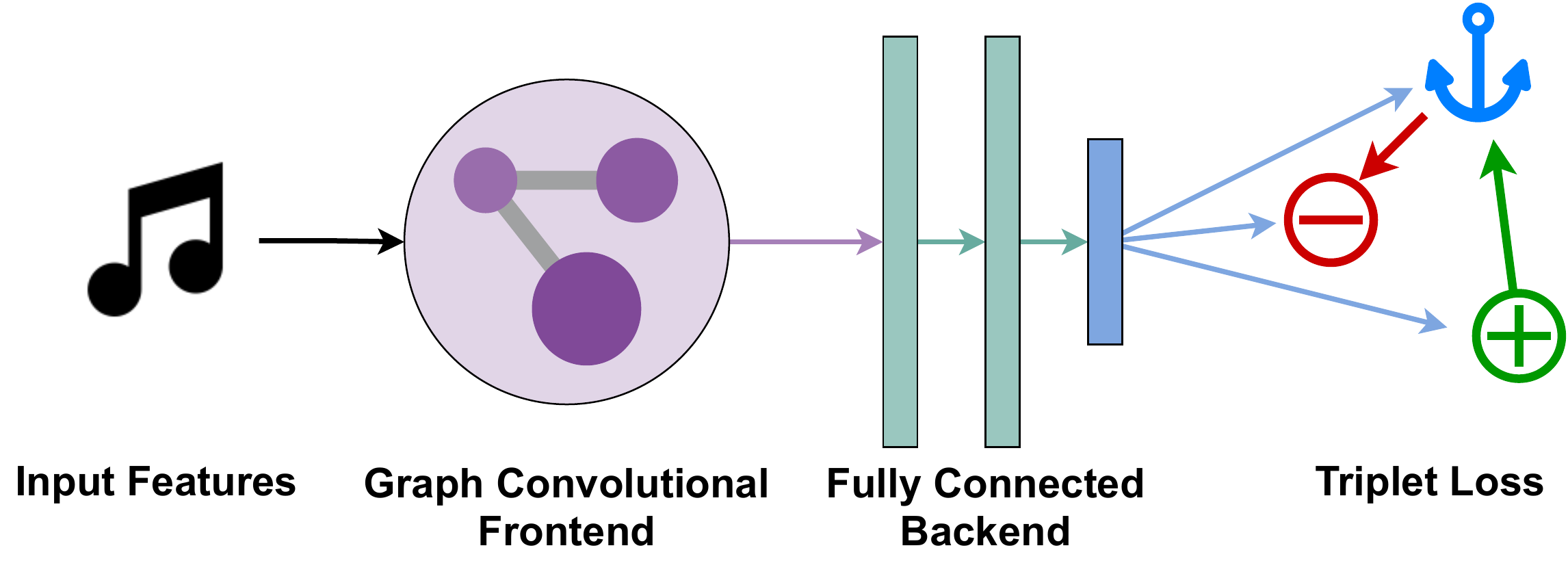}}
 \caption{Overview of the graph neural network we use in this paper. First, the input
 features $\mb{x}_v$ are first passed through a \emph{front-end} of graph convolution layers (see Sec.~\ref{sec:graph_convolutions}
 for details); then, the output of the front-end is passed through a traditional deep neural network \emph{back-end} to
 compute the final embeddings $\mb{y}_v$ of artist nodes. 
 Based on these embeddings, we use the triplet loss to train the network
 to project similar artists (positive, green) closer to the anchor, and dissimilar ones (negative, red) further away.
 }
 \label{fig:gnn_model}
\end{figure}

The GNN we use in this paper comprises two parts: first, a block of
\emph{graph convolutions} (GC) processes each node's features and combines them
with the features of adjacent nodes; then, another block of fully connected
layers project the resulting feature representation into the target embedding
space. See Fig.~\ref{fig:gnn_model} for an overview.

We train the model using the triplet loss, in an identical setup as
the baseline model. Viewing the proposed GNN from this angle, the only difference
of the GNN from a standard embedding network is the additional 
\emph{Graph Convolutional Frontend}. In other words, if we remove all graph
convolutional layers, we arrive at our baseline model, a fully connected Deep Neural Network (DNN).

\subsubsection{Graph Convolutions}\label{sec:graph_convolutions}

The graph convolution algorithm, as defined in~\cite{hamilton_inductiverepresentationlearning_2017,ying_graphconvolutionalneural_2018},
features two operations which are not found in classic neural networks: a \emph{neighborhood function}
\(\mathcal{N}(\cdot)\), which yields the set of neighbors of a given node; and an \emph{aggregation function}, which computes a vector-valued aggregation of a set of input vectors.

As a neighborhood function, most models use guided or uniform sub-sampling of
the graph structure~\cite{oh_advancinggraphsagedatadriven_2019a,hamilton_inductiverepresentationlearning_2017,ying_graphconvolutionalneural_2018}.
This limits the number of neighbors to be processed for each node, and is often necessary to 
adhere to computational limits. As aggregation functions, models commonly apply pooling operators, LSTM networks, 
or (weighted) point-wise averages~\cite{hamilton_inductiverepresentationlearning_2017}.

In this work, we take a simple approach, and use point-wise weighted averaging to aggregate
neighbor representations, and select the strongest 25 connections as neighbors (if weights are not
available, we use the simple average of random 25 connections). This enables us to
use a single sparse dot-product with an adjacency matrix to select and aggregate neighborhood 
embeddings. Note that this is not the full adjacency matrix of the complete graph, as we select only the parts of the
graph which are necessary for computing embeddings for the nodes in a mini-batch.

Algorithm~\ref{alg:graph_convolution} describes the inner workings of the graph convolution
block of our model. Here, the matrix \( \mb{X} \in \mathbb{R}^{D \times V} \) stores the \(D\)-dimensional
features of all \(V\) nodes, the symmetric sparse matrix \( \mb{A} \in \mathbb{R}^{V \times V} \)
defines the connectivity of the graph, and \( \mathcal{N}(v) \) is a neighborhood function
which returns all connected nodes of a given node \(v\) (here, all non-zero elements in the \(v\)\textsuperscript{th}
row of \(\mb{A}\)). 

\begin{figure}
 \centerline{
    \includegraphics{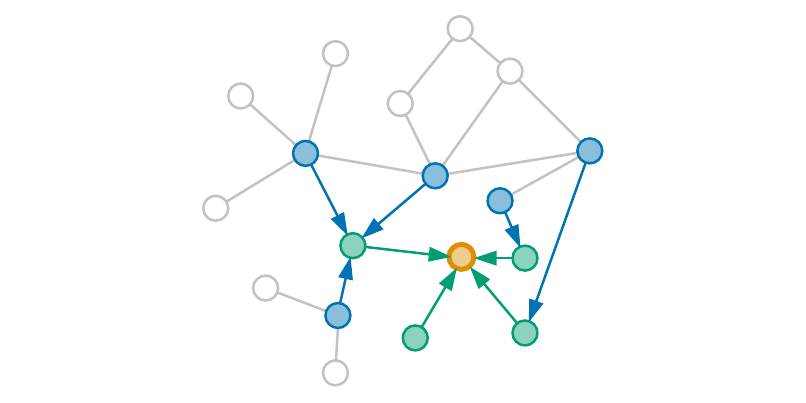}}
 \caption{
    Tracing the graph to find the necessary input nodes for embedding the 
    target node (orange). Each graph convolution layer requires tracing one
    step in the graph. Here, we show the trace for a stack of two such layers.
    To compute the embedding of the target node in the last layer, we need the representations
    from the previous layer of itself and its neighbors (green). In turn, to compute
    these representations, we need to expand the neighborhood by one additional step in the
    preceding GC layer (blue).
    Thus, the features of all colored nodes must be fed to the first graph convolution layer.
}
\label{fig:graph_trace}
\end{figure}

To compute the output of a graph convolution layer for a node, we need to know its neighbors. Therefore, to
compute the embeddings for a mini-batch of nodes \(\mathcal{V}\), we need to know which nodes are in their
joint neighborhood. Thus, before the actual processing, we first need to trace the graph to find the node
features necessary to compute the embeddings of the nodes in the mini-batch. This is shown in Fig.~\ref{fig:graph_trace},
and formalized in lines 1--4 of Alg.~\ref{alg:graph_convolution}.

\begin{algorithm}[h!]
\caption{\textsc{graph convolution block}}
\label{alg:graph_convolution}
\SetKwInOut{Input}{Input}
\SetKwInOut{Output}{Output}
\SetKwComment{Comment}{\(\triangleright\)\ }{}
\newcommand\mycommfont[1]{\small\rmfamily\textcolor{gray}{#1}}
\SetCommentSty{mycommfont}

\Input{Node input features \( \mb{X} \).\linebreak
    Sparse connectivity matrix \( \mb{A} \). \linebreak
    Nodes in mini-batch \( \mathcal{V} \subset [1\ldots V] \).
    
}
\Output{Node output representation \(\mb{X}_K\)
}
    \BlankLine
    \Comment{Trace back input nodes for each layer.}
    \( \mathcal{V}_K \leftarrow \mathcal{V} \) \;
    \For{\(k=K-1 \ldots 0\)}{
        \(\mathcal{V}_k \leftarrow \bigcup_{v \in \mathcal{V}_{k + 1}}{\mathcal{N}(v)} \) \;
    }
    \Comment{Select input features for first layer. We use \(\mathbf{M}[r,c]\) to denote selecting
             \(r\) rows and \(c\) columns from a matrix $\mathbf{M}$.}
    \( \mb{X}_0 = \mb{X}\left[\cdot, \mathcal{V}_0\right] \) \;
    \For{$k=1 \ldots K$}{
        \( \mb{A}_k = \mb{A}\left[\mathcal{V}_{k-1}, \mathcal{V}_k\right] \) \;
        \( \mb{N}_k = \sigma\left(\mb{Q}_k \cdot \mb{X}_{k-1} \right) \cdot \mb{A}_k \) \;
        \( \mb{X}_k \leftarrow \sigma\left(\mb{W}_k \cdot \begin{bmatrix} \mb{N}_k \\ \mb{X}_{k-1}\left[\cdot, \mathcal{V}_k\right] \end{bmatrix} \right) \) \;
        \Comment{\(l_2\)-normalize embeddings of each output node.}
        \( \mb{X}_k \leftarrow \begin{bmatrix} \frac{\mb{x}_v}{\| \mb{x}_v \|_2} \mid v \in \mathcal{V}_k \end{bmatrix} \) \;
    }
    \Return{\(\mb{X}_K\)}
\end{algorithm}

At the core of each graph
convolution layer \(k \in [1\ldots K]\) there are two non-linear projections, parameterized by 
projection matrices \(\mb{Q}_k \in \mathbb{R}^{H_{Q_k} \times D}\) and \(\mb{W}_k \in \mathbb{R}^{H_{W_k} \times (H_Q + D)}\), and a point-wise non-linear activation
function \(\sigma\), in our case, the Exponential Linear Unit function (ELU). Here, \(H_{Q_k}\) and \(H_{W_k}\) are the output dimensions of the respective projections. 
The last output, \( \mb{X}_K \in \mathbb{R}^{H_{W_K} \times V}\), holds the \(l_2\)-normalized representations of each node in the 
mini-batch in its columns. It is fed into the following fully connected layers, which then compute the output embedding \(\mb{y}_v\) of a node. Finally, these embeddings are used to compute the triplet loss and back-propagate it through the GNN.

\section{Datasets}\label{sec:datasets}

Many published studies on the topic of artist similarity are limited by data: datasets including artists, their similarity relations, and their features comprise at most hundreds to a few thousand artists. 
In addition, the quality of the ground truth provided is often based on 3\textsuperscript{rd} party APIs with obscure similarity methods like the last.fm API, rather than based on data curated by human experts. 

For instance, in~\cite{oramas_semanticbasedapproachartist_2015}, two datasets are provided, one with \textasciitilde 2k artists and similarity based on last.fm relations, and another with only 268 artists, but based on relations curated by human experts.  In~\cite{schedl_harvestingmicroblogscontextual_2014}, a dataset of 1,677 artists based on last.fm similarity relations is used for evaluation. Also, the dataset used in the Audio Music Similarity and Retrieval (AMS) MIREX task, which was manually curated, contains data about only 602 artists. Other works, like~\cite{lee_disentangledmultidimensionalmetric_2020}, use tag data shared among tracks or artists as a proxy for similarity estimation---which can be considered as a weak signal of similarity---and use a small set of 879 human-labeled triplets for evaluation.

For all these issues regarding existing datasets, we compiled a new dataset, the \dataset{} Dataset, which we describe in the following.

\subsection{The \dataset{} Dataset}

For the \dataset{} (``\textbf{O}h, what a \textbf{L}arge \textbf{G}raph of \textbf{A}rtists'') dataset,
we bring together content-based low-level features from AcousticBrainz~\cite{porter_acousticbrainzcommunityplatform_2015}, and similarity
relations from AllMusic.
Assembling the data works as follows:
\begin{enumerate}
    \item Select a common pool of artists based on the unique artists in the Million Song Dataset~\cite{bertin-mahieux_millionsongdataset_2011}.
    \item Map the available MusicBrainz IDs of the artists to AllMusic IDs using mapping available from \mbox{MusicBrainz}.
    \item For each artist, obtain the list of ``related'' artists from AllMusic; this data can be licensed and accessed on their website. 
          Use only related artists which can be mapped back to MusicBrainz.
    \item Using MusicBrainz, select up to 25 tracks for each artist using their API, and collect the low-level features
          of the tracks from AcousticBrainz.
    \item Compute the track feature centroid of each artist.
\end{enumerate}

In total, the dataset comprises 17,673 artists connected by 101,029 similarity relations. On average,
each artist is connected to 11.43 other artists. The quartiles are at 3, 7, and 16 connections per artist.
The lower 10\% of artists have only one connection, the top 10\% have at least 27. 

While the dataset size is still small compared to industrial catalog sizes, it is significantly bigger
than other datasets available for this task. Its size and available features will allow us to apply
more data-driven machine learning methods to the problem of artist similarity.\footnote{The procedure to 
assemble the dataset, including relevant metadata, is available on \mbox{\url{https://gitlab.com/fdlm/olga/}}.}

For our experiments, we partition the artists following an 80/10/10 split into
14,139 training, 1767 validation, and 1767 test artists.

\subsection{Proprietary Dataset}

We also use a larger proprietary dataset to demonstrate the scalability of our approach.
Here, explicit feedback from listeners of a 
music streaming service is used to define whether two artists are similar or not. 

For artist features, we use the centroid of an artist's track features. These
track features are \emph{musicological} attributes annotated by experts, and
comprise hundreds of content-based characteristics such as
``amount of electric guitar'', or ``prevalence of groove''.

In total, this dataset consists of 136,731 artists connected by 3,277,677
similarity relations. The number of connections per artists is a top-heavy
distribution with few artists sharing most of the connections: the top 10\%
are each connected to more than 134 others, while the bottom 10\% to only one.
The quartiles are at 2, 5, and 48 connections per artist.

We follow the same partition strategy as for the \dataset{} dataset, which results in
109,383 training, 13,674 validation, and 13,674 test artists.

\section{Experiments}\label{sec:experiments}

Our experiments aim to evaluate how well the embeddings produced by our model
capture artist similarity. To this end, we set up a ranking
scenario:
given an artist, we collect its \(K\) nearest neighbors sorted by ascending 
distance, and evaluate the quality of this ranking. To quantify this, we use normalized 
discounted cumulative gain~\cite{jarvelin_cumulatedgainbasedevaluation_2002} with
a high cut-off at \(K=200\) (``ndcg@200''). We prefer this metric over others, because it was shown
that at high cut-off values, it provides better discriminative power, as well as robustness to sparsity
bias (and, to moderate degree, popularity bias)~\cite{valcarce_robustnessdiscriminativepower_2018}. 
Formally, given an artist \(a\) with an ideal list of similar artists \(\mathbf{s}\) (sorted by relevance), 
the \(\operatorname{nDCG}_K\) of a predicted list of similar artists \(\mathbf{\hat{s}}\) is defined as:
\[
\operatorname{nDCG_K}(a, \mathbf{\hat{s}}, \mathbf{s}) = 
    \frac{\sum_{k=1}^{K} g(\hat{s}_k, a) d(k)}
         {\sum_{k=1}^{K} g(s_k, a) d(k)},
\]
where \(g(\cdot, a)\), the \emph{gain}, is 1 if an artist is indeed similar to \(a\), and 0 otherwise, and 
\( d(k) = \log_2^{-1}(k + 1) \)
the \emph{discounting} factor, weights top rankings higher than the tail of the list.

In the following, we first explain the models, their training details, the features,
and the evaluation data used in our experiments. Then, we show, compare and analyze
the results.

\subsection{Models}

As explained in Sec.~\ref{sec:model_overview}, a GNN with
no graph convolutional layers is identical to our baseline model (i.e. a DNN trained
using triplet loss). This allows us to fixate hyper-parameters between baseline and
the proposed GNN, and isolate the effect of adding graph convolutions to the model.
For each dataset, we thus train and evaluate four models with~0~to~3 graph convolutional layers.

The other hyper-parameters remain fixed: each layers in the graph 
convolutional front-end consists of 256 ELUs~\cite{clevert_fastaccuratedeep_2016};
the back-end comprises two layers of 256 ELUs each, and one linear output 
layer with a 100 dimensions; we train the networks using the ADAM optimizer~\cite{kingma_adammethodstochastic_2015} 
with a linear learning-rate warm-up~\cite{ma_adequacyuntunedwarmup_2021} for the first epoch, and following a cosine learning rate decay~\cite{loshchilov_sgdrstochasticgradient_2017b} for the remaining 49 epochs
(in contrast to~\cite{loshchilov_sgdrstochasticgradient_2017b}, we do not use warm-restarts); for selecting triplets,
we apply distance-weighted sampling~\cite{wu_samplingmattersdeep_2017}, and use a margin of \( \Delta = 0.2 \) in the loss; 
finally, as distance measure, we use Euclidean distance between \(l_2\)-normalized embeddings.

We are able to train the largest model with 3 graph convolutional layers within 2 hours on the proprietary dataset, and under
5 minutes on \dataset{}, using a Tesla P100 GPU and 8 CPU threads for data loading. 

\subsection{Features}

We build artist-level features by averaging track-level features of the artist's tracks. 
Depending on the dataset, we have different types of features at hand. 

In the \dataset{} dataset,
we 
have low-level audio features as extracted by the Essentia library.\footnote{See \url{https://essentia.upf.edu/streaming_extractor_music.html#music-descriptors}} 
These features represent track-level statistics about the loudness, dynamics and spectral shape of the signal, but they also
include more abstract descriptors of rhythm and tonal information, such as bpm and the average pitch class profile.
We select all numeric features and pre-process them as follows: we apply element-wise standardization, discard
features with missing values, and flatten all numbers into a single vector of 2613 elements.

In the proprietary dataset, we use numeric musicological descriptors annotated by
experts (for example, ``the nasality of the singing voice''). We apply the same pre-processing 
for these, resulting in a total of 170 values.

Using two different types of content features gives us the opportunity to evaluate
the utility of our graph model under different circumstances, or more precisely,
features of different quality and signal-to-noise ratio. The low-level audio-based features
available in the \dataset{} dataset are undoubtedly noisier and less specific than the high-level
musical descriptors manually annotated by experts, which are available in the proprietary dataset. 
Experimenting with both permits us to gauge the effect of using the graph topology for different data
representations.

In addition, we also train models with \emph{random vectors} as features.
For each artist, we uniformly sample a random vector of the same dimension as the real features, and 
and keep it constant throughout training and testing. 
This way, we can differentiate between the performance of the real features and the performance of using the graph
topology in the model: the results of a model with no graph convolutions is only due
to the features, while the results of a model with graph convolutions but random
features is only due to the usage of the graph topology.

\subsection{Evaluation Data}

As described in Section~\ref{sec:datasets}, we partition artists into a training, validation
and test set. When evaluating on the validation or test sets, we only consider
artists from these sets as candidates and potential true positives. Specifically,
let \( \mathcal{V}_{\mathrm{eval}} \) be the set of evaluation artists, we only compute embeddings
for those, and retrieve nearest neighbors from this set, and only consider ground truth
similarity connections within~\( \mathcal{V}_{\mathrm{eval}} \).

\begin{figure}
\centerline{
    \includegraphics[width=\columnwidth]{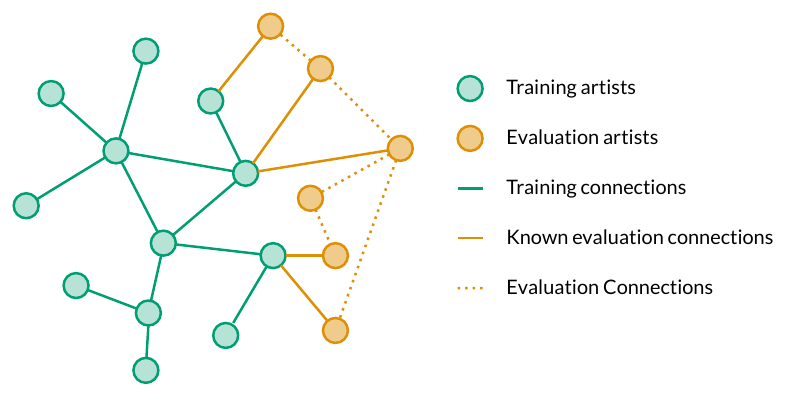}
}
\caption{
    Artist nodes and their connections used for training (green) and evaluation (orange).
    During training, only green nodes and connections are used. When evaluating, we extend
    the graph with the orange nodes, but only add connections between validation and training artists.
    Connections among evaluation artists (dotted orange) remain hidden. We then compute the embeddings
    of all evaluation artists, and evaluate based on the hidden evaluation connections.
}
\label{fig:eval_data}
\end{figure}

This notion is more nuanced in the case of GNNs. Here, we want to exploit the
\emph{known artist graph topology} (i.e.,\ which artists are connected to each other)
when computing the embeddings. To this end, we use all connections between artists in
\( \mathcal{V}_{\mathrm{train}} \) (the training set) \emph{and} connections between
artists in \( \mathcal{V}_{\mathrm{train}} \) and \( \mathcal{V}_{\mathrm{eval}} \). 
This process is outlined in Fig.~\ref{fig:eval_data}.

Note that this does not leak information between train and evaluation sets; the features
of evaluation artists have not been seen during training, and connections within
the evaluation set---these are the ones we want to predict---remain hidden. 

\subsection{Results}

Table~\ref{tab:results_table} compares the baseline model with the proposed GNN. 
We can see that the GNN easily out-performs the DNN. It achieves an NDCG@200 of 0.55 vs.\ 0.24
on the \dataset{} dataset, and 0.57 vs.\ 0.44 on the proprietary dataset. The table also demonstrates that the
graph topology is more predictive of artist similarity than content-based features: the GNN, 
using random features, achieves better results than a DNN using informative features for
both datasets (0.45 vs.\ 0.24 on \dataset{}, and 0.52 vs 0.44 on the proprietary dataset). 

\begin{table}[]
\centering
\begin{tabular}{llcc}
\toprule
\textbf{Dataset}             & \textbf{Features} & \textbf{DNN} & \textbf{GNN} \\ \midrule
\multirow{2}{*}{OLGA}        & Random           & 0.02         & 0.45         \\
                             & AcousticBrainz   & 0.24         & 0.55         \\ \midrule
\multirow{2}{*}{Proprietary} & Random           & 0.00         & 0.52         \\
                             & Musicological    & 0.44         & 0.57         \\ \bottomrule
\end{tabular}

\caption{
NDCG@200 for the baseline (DNN) and the proposed model with 3 graph convolution layers (GNN), 
using features or random vectors as input. The GNN with real features as input gives the
best results. Most strikingly, the GNN with random features---using only the known
graph topology---out-performs the baseline DNN with informative features.}
\label{tab:results_table}
\end{table}

Additionally, the results indicate---perhaps to little surprise---that low-level audio
features in the \dataset{} dataset are less informative than manually annotated high-level
features in the proprietary dataset. Although the proprietary dataset poses a more difficult
challenge due to the much larger number of candidates (14k vs.\ 1.8k), the DNN---which can only use
the features---improves more over the random baseline in the proprietary dataset (+0.44), compared to 
the improvement (+0.22) on \dataset{}. These are only indications; for a definitive analysis,
we would need to use the exact same features in both datasets.

Similarly, we could argue that the topology in the proprietary dataset seems more coherent
than in the \dataset{} dataset. We can judge this by observing the performance gain obtained
by a GNN with random feature---which can only leverage the
graph topology to find similar artists---compared to a
completely random baseline (random features without GC layers).
In the proprietary dataset, this performance gain is +0.52, while in the \dataset{} dataset,
only +0.43. Again, while this is not a definitive analysis (other factors may play a
role), it indicates that the large amounts of user feedback used to generate ground truth 
in the proprietary dataset give stable and high-quality similarity connections.

\begin{figure}
\begin{subfigure}{\columnwidth}
    \centerline{
        \includegraphics[width=\columnwidth]{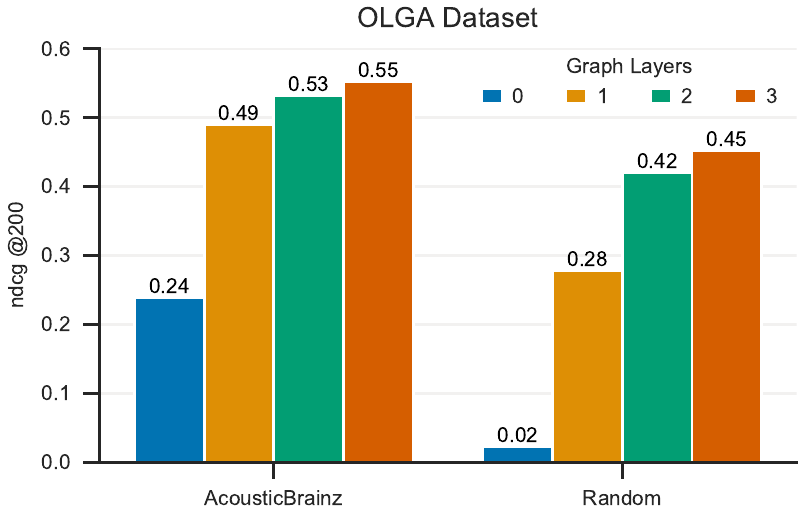}
    }
\end{subfigure}
    \vskip1em
\begin{subfigure}{\columnwidth}
    \centerline{
        \includegraphics[width=\columnwidth]{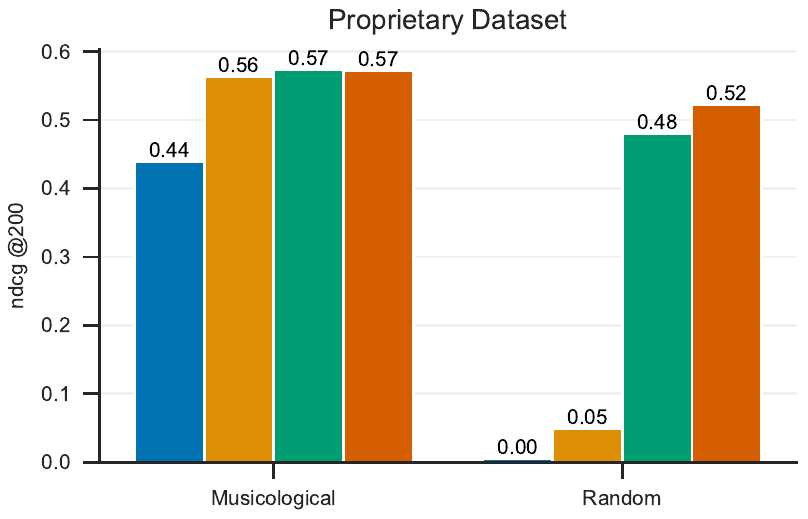}
    }
\end{subfigure}
\caption{Results on the \dataset{} (top) and the proprietary dataset (bottom) with different numbers
         of graph convolution layers, using either the given features (left) or random
         vectors as features (right).}
\label{fig:results_figure}
\end{figure}

Figure~\ref{fig:results_figure} depicts the results for each model
and feature set depending on the number of graph convolutional layers used. (Recall that
a GNN with 0 graph convolutions corresponds to the baseline DNN.) 
In the \dataset{} dataset, we see the scores
increase with every added layer. This effect is less pronounced in the proprietary dataset,
where adding graph convolutions does help significantly, but results plateau after the
first graph convolutional layer. We believe this is due to the quality and informativeness
of the features: the low-level features in the \dataset{} dataset provide less information
about artist similarity than high-level expertly annotated musicological attributes in 
the proprietary dataset. Therefore, exploiting contextual information through 
graph convolutions results in more uplift in the \dataset{} dataset than in the proprietary
one.

Looking at the scores obtained using random features (where the model depends solely
on exploiting the graph topology), we observe two remarkable results. First, whereas
one graph convolutional layer suffices to out-perform the feature-based baseline in the
\dataset{} dataset (0.28 vs.\ 0.24), using only one GC layer does not produce meaningful results
(0.05) in the proprietary dataset. We believe this is due to the different sizes of the 
respective test sets: 14k in the proprietary dataset, while only 1.8k in \dataset{}. Using only
a very local context seems to be enough to meaningfully organize the artists in a smaller
dataset. 

Second, most performance gains are obtained with two GC layers, while adding
the third GC layer pushes the results to a much lesser degree. 
Our explanation for this effect is that most similar artists are connected through at
least one other, common artist. In other words, most artists form similarity cliques 
with at least two other artists. Within these cliques, in which every artist is connected
to all others, missing connections are easily retrieved by no more than 2 graph convolutions.

In fact, in the \dataset{} dataset, 
\textasciitilde71\% of all cliques fulfill this requirement. This means that, for any hidden
similarity link in the data, in 71\% of cases, the true similar artist is within 2 steps in the
graph---which corresponds to using two GC layers.

\section{Summary and Future Work}\label{sec:summary}

In this paper, we described a hybrid approach to computing artist similarity, which uses
graph neural networks to combine content-based features with explicit relations between
artists. To evaluate our approach, we assembled a novel academic dataset with
17,673 artists, their features, and their similarity relations. Additionally, we used
a much larger proprietary dataset to show the scalability of our method. The results showed
that leveraging known similarity relations between artists can be more effective for
understanding their similarity than high-quality features, and that combining
both gives the best results.

Our work is a first step towards models that directly use known relations between
musical entities---like tracks, artists, or even genres---or even across these
modalities. Multi-modal connections could also help predicting artist similarity; we
could add collaborations, or band membership connections to the graph. 
Finally, it would be interesting to analyze  the effect of our approach on long-tail recommendations and/or the cold-start problem.

\bibliography{ismir2021}

\end{document}